\documentclass[11pt,twoside]{article}
\usepackage{asp2004}
\usepackage{psfig}
\usepackage{epsf}
\usepackage{graphics}
\usepackage{lscape}
\newcommand{\HI}{\mbox{H~{\sc i}}}

\def\edcomment#1{\iffalse\marginpar{\raggedright\sl#1\/}\else\relax\fi}
\reversemarginpar

\begin{document}
\title{High velocity HI in NGC 6946 and extra-planar gas in NGC 253}
\author{R. Boomsma}
\affil{Kapteyn Astronomical Institute, P.O. Box 800, 9700 AV Groningen, the Netherlands}
\author{T.A. Oosterloo}
\affil{ASTRON, P.O. Box 2, 7990 AA Dwingeloo, the Netherlands}
\author{F. Fraternali}
\affil{Theoretical Physics, University of Oxford, 1 Keble Road, Oxford OX1 3NP, UK}
\author{J.M. van der Hulst}
\affil{Kapteyn Astronomical Institute, Groningen, the Netherlands}
\author{R. Sancisi}
\affil{INAF-Osservatorio Astronomico di Bologna, via Ranzani 1, I-40127 Bologna, Italy\\\& Kapteyn Astronomical Institute, Groningen, the Netherlands}

\begin{abstract}
Multi-wavelength observations of nearby spiral galaxies have shown
that neutral and ionized gas are present up to a few kpc from the disk
and that star formation and supernovae probably play an important role
in bringing gas into the halo. We have obtained very sensitive \HI\
observations of the face-on galaxy \mbox{NGC 6946} and of the nearly
edge-on starburst galaxy \mbox{NGC 253}. We find high velocity \HI\
clouds in \mbox{NGC 6946} and extra-planar gas with anomalous
velocities in \mbox{NGC 253}. In both galaxies there seems to be a
close connection between the star-forming disk and the halo \HI. In
the outer parts of \mbox{NGC 6946} there is also evidence for recent
gas accretion.
\end{abstract}
\thispagestyle{plain}

\section{Introduction}
\HI\ has been detected in the halos of several spiral galaxies. Studies of the edge-on galaxies \mbox{NGC 891} \citep{swa97, fra03} and \mbox{UGC 7321} \citep{mat03}, and the intermediate inclination galaxy \mbox{NGC 2403} \citep{sch00,fra03} have revealed a thick layer of \HI\ rotating more slowely than the underlying disk (see Oosterloo et al., this conference, for an overview of these studies). The origin of such a thick layer or halo is unknown. There are various possibilities. One of them is the galactic fountain mechanism according to which active star formation in the disk and supernovae are responsible for blowing the gas into the halo.\\
 The galaxies mentioned above are seen under suitable orientations for a study of the vertical \HI\ distribution and the general kinematics. Vertical motions are instead best observed in face-on galaxies.\\
 For this purpose we have obtained new, deep \HI\ observations with the Westerbork Synthesis Radio Telescope (WSRT) of \mbox{NGC 6946}, a nearby face-on spiral galaxy showing moderate star formation activity.\\
In addition we have chosen \mbox{NGC 253} as a case of a galaxy with an extreme starburst, in order to trace the effects on the \HI\ in such a limiting case. For this galaxy we have used \HI\ data from the Australia Telescope Compact Array (ATCA).\\

\section{NGC 6946}

\subsection{Holes and high velocity gas}
\mbox{NGC 6946} is a nearby (6 Mpc) face-on spiral galaxy of type Scd. It has a moderate star formation activity similar to that of the Milky Way. The \HI\ disk shows a large number of holes (see Fig. 1, left panel). The holes may have originated from the explosions of many supernovae.\\
\begin{figure}[!ht]
\plotone{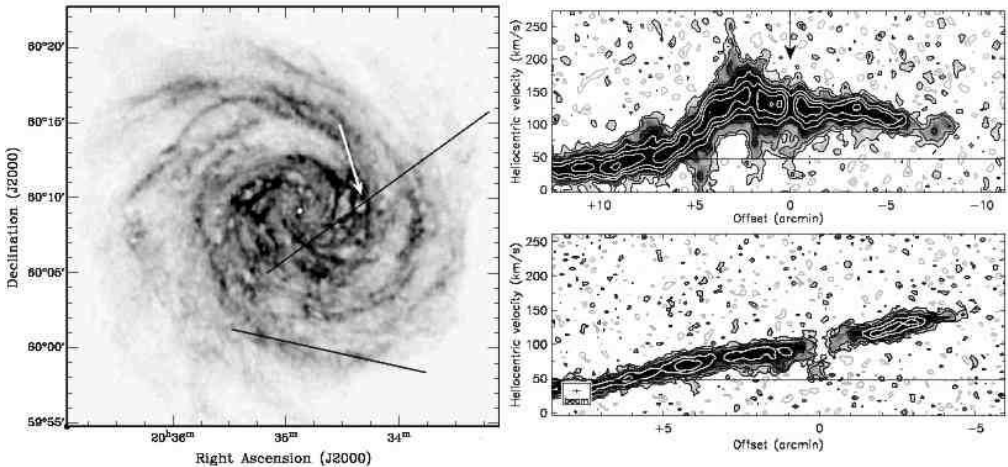}
\caption{The left panel shows the total \HI\ distribution in \mbox{NGC 6946}. The two lines indicate the locations of the cuts corresponding to the position-velocity diagrams on the right. The black arrow in the top right panel and the white arrow in the left panel point at the position of an HI hole. The horizontal lines in the position velocity diagrams show the systemic velocity of \mbox{NGC 6946} (48 km/s).}
\end{figure}
Besides the large number of holes, \mbox{NGC 6946} shows a significant amount of \HI\ with anomalous velocities. \citet{kam93} showed that \mbox{NGC 6946} has a widespread high velocity gas component, but they could only reveal it by heavily smoothing the data.\\ With our observations we have the sensitivity to study the gas distribution and kinematics in detail. As shown in the top right panel of Fig. 1, we are detecting several individual complexes of \HI\ with anomalous velocities. Most of the complexes appear to have velocities lower than rotation, such as the 'beard' gas in \mbox{NGC 2403} \citep{fra01}. But there are also velocities higher than the local disk rotational velocity, sometimes in excess of 100 km/s. The inclination of \mbox{NGC 6946} is such that these are probably vertical motions.\\
 The arrows in the left and top right panels of Fig. 1 indicate the position of a big \HI\ hole. Some anomalous \HI\ is seen in this same direction on the low velocity side, but it is not clear whether this gas is originating from the hole.\\
 This slice shows that all the anomalous \HI\ complexes are in the direction of the inner bright optical disk (R$_{25}=5.75^{\prime}$), while at larger radii the gas disk seems to be more quiet. This supports the galactic fountain hypothesis according to which massive star formation and supernovae are responsible for blowing the gas out of the disk. To further illustrate this, we have constructed a map of the anomalous \HI, by integrating over velocities that differ more than a certain amount from normal galactic rotation. The result is shown in Fig. 2. We find about $1.3\times10^{8}$ M$_{\odot}$ of \HI\ with velocities deviating more than $\pm50$ km/s from the local rotation. This is about 2\% of the total \HI\ mass of \mbox{NGC 6946} ( $6\times10^{9}$ M$_{\odot}$).\\ Clearly, nearly all the gas with velocities higher than 50 km/s is found in the direction of the bright optical disk (which corresponds to the bright H$\alpha$ distribution shown in the left panel of Fig. 2). This strongly suggests that star formation and high supernova rate are at the origin of the anomalous \HI\ and that the galactic fountain mechanism is responsible.\\

 However, Fig. 2 shows that some anomalous \HI\ is also found at the outer edges of the gas disk where there is little star formation going on. For this gas, therefore, it seems that a different explanation is needed.\\

\begin{figure}[!ht]
\plotone{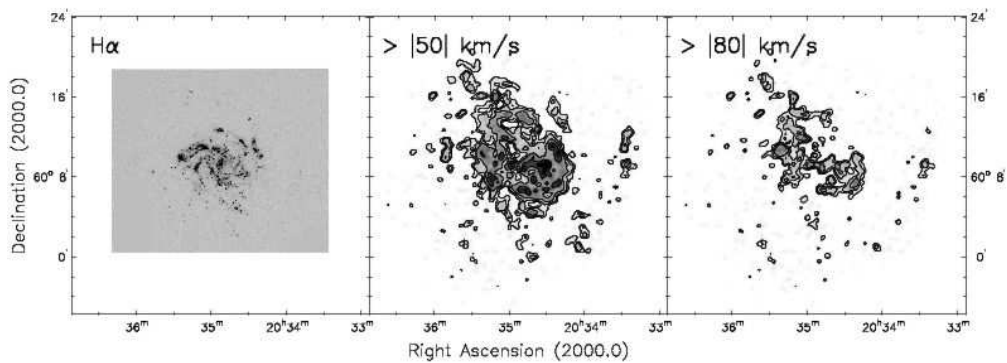}
\caption{The left panel shows the H$\alpha$ emission from NGC 6946 \citep[][]{fer98}. The middle and right panel show the anomalous \HI\ with velocities deviating more than $\pm50$ and $\pm80$ km/s from galactic rotation.}
\end{figure}

\subsection{Accretion}

The middle and right panels of Fig. 2 show a few distinct clouds in the outer parts of the \HI\ disk of \mbox{NGC 6946}. One of them is associated with a hole in the outer disk (Fig 1, bottom right panel). In this case, star formation is unlikely to be the cause. This is reminiscent of the \HI\ hole and gas clouds found in M101 by \citet{vdh88}. Perhaps, as suggested for the M101 hole, also here a collision with a small companion galaxy or a dense gas cloud may have occurred.\\
 Other indications of accretion in \mbox{NGC 6946} are noticable at low resolution. In Fig. 3 the disk shows a large plume at the northwestern side of the disk. The gas in this plume has the same velocity as the two companions at the same side of \mbox{NGC 6946}. A follow-up observation with the WSRT centered at a position between \mbox{NGC 6946} and the companions has not revealed any bridge between them. Also the gas kinematics in the two dwarf galaxies seems unperturbed. Therefore the gas in the plume is probably not originating from these companions. Perhaps we are witnessing the accretion of a third companion. Accretion phenomena seem to be common in galaxies (Van der Hulst \& Sancisi, paper at this symposium). An example is the Sagittarius Dwarf in the Milky Way \citep{iba99}.

\begin{figure}[!ht]
\plotone{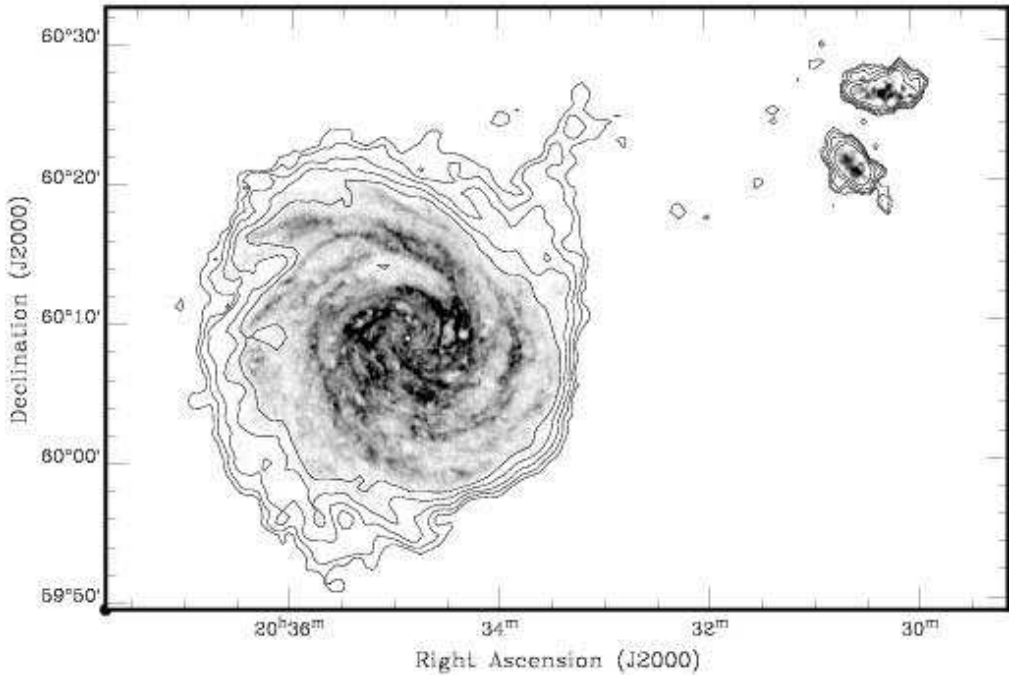}
\caption{The \HI\ ditribution in NGC 6946 and its companions at a resolution of 13$^{\prime\prime}$ (grayscale) and 60$^{\prime\prime}$ (contours). Contours are at 1.3, 2.6, 5.2, 10.4, and 20.8$\times 10^{19}$ cm$^{-2}$.}
\end{figure}

\section{NGC 253}

\mbox{NGC 253} is a barred Sc galaxy in the Sculptor group of
galaxies, at a distance of 3.9 Mpc. It is one of best nearby examples
of a nuclear starburst galaxy.\\ The high star formation rate in the
inner parts of the disk makes \mbox{NGC 253} an interesting case for
the study of the galactic fountain phenomenon. The massive star
formation and the related high supernova rate should blow a significant
amount of gas out of the disk.\\
\begin{figure}[!ht]
\plottwo{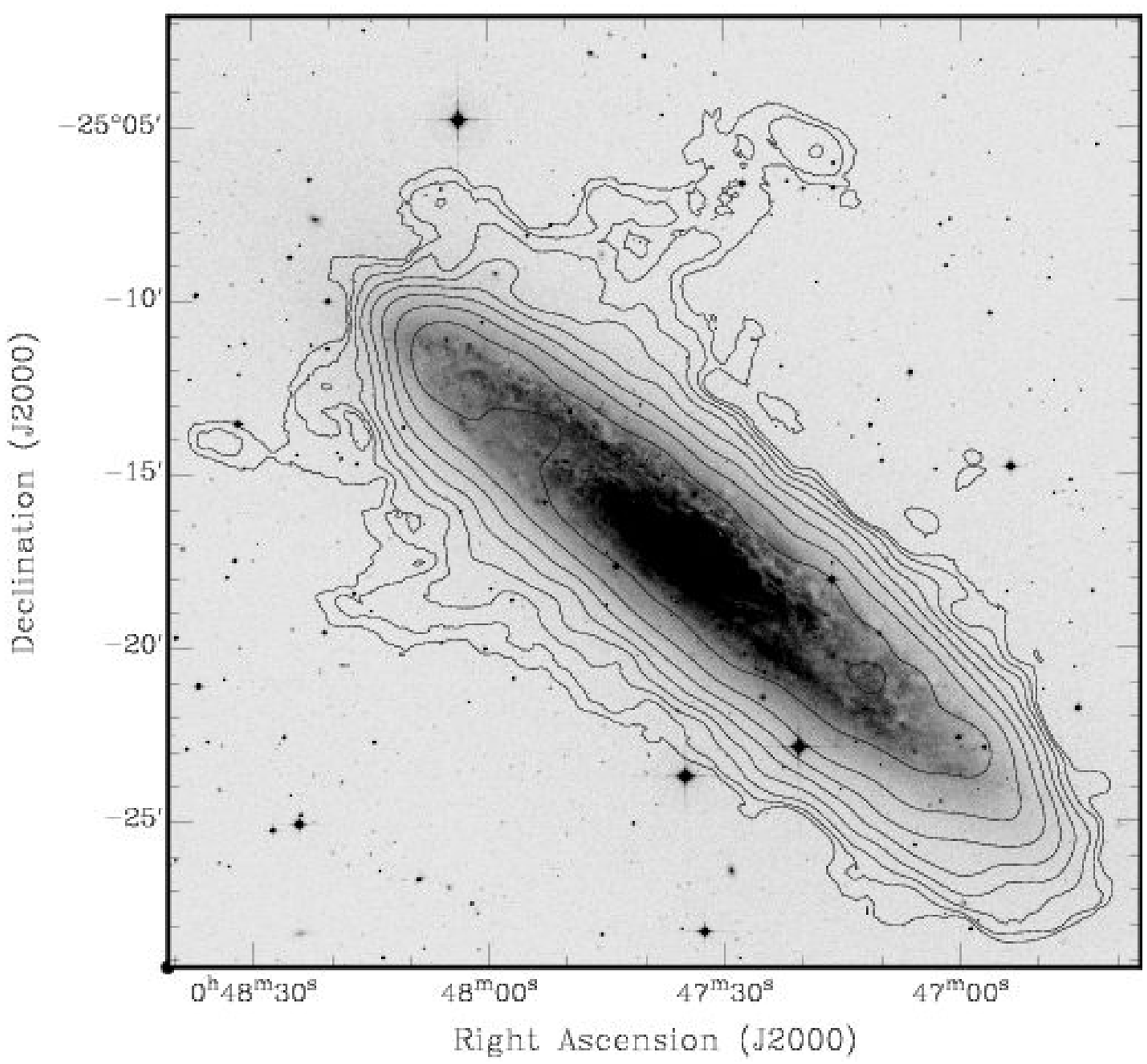}{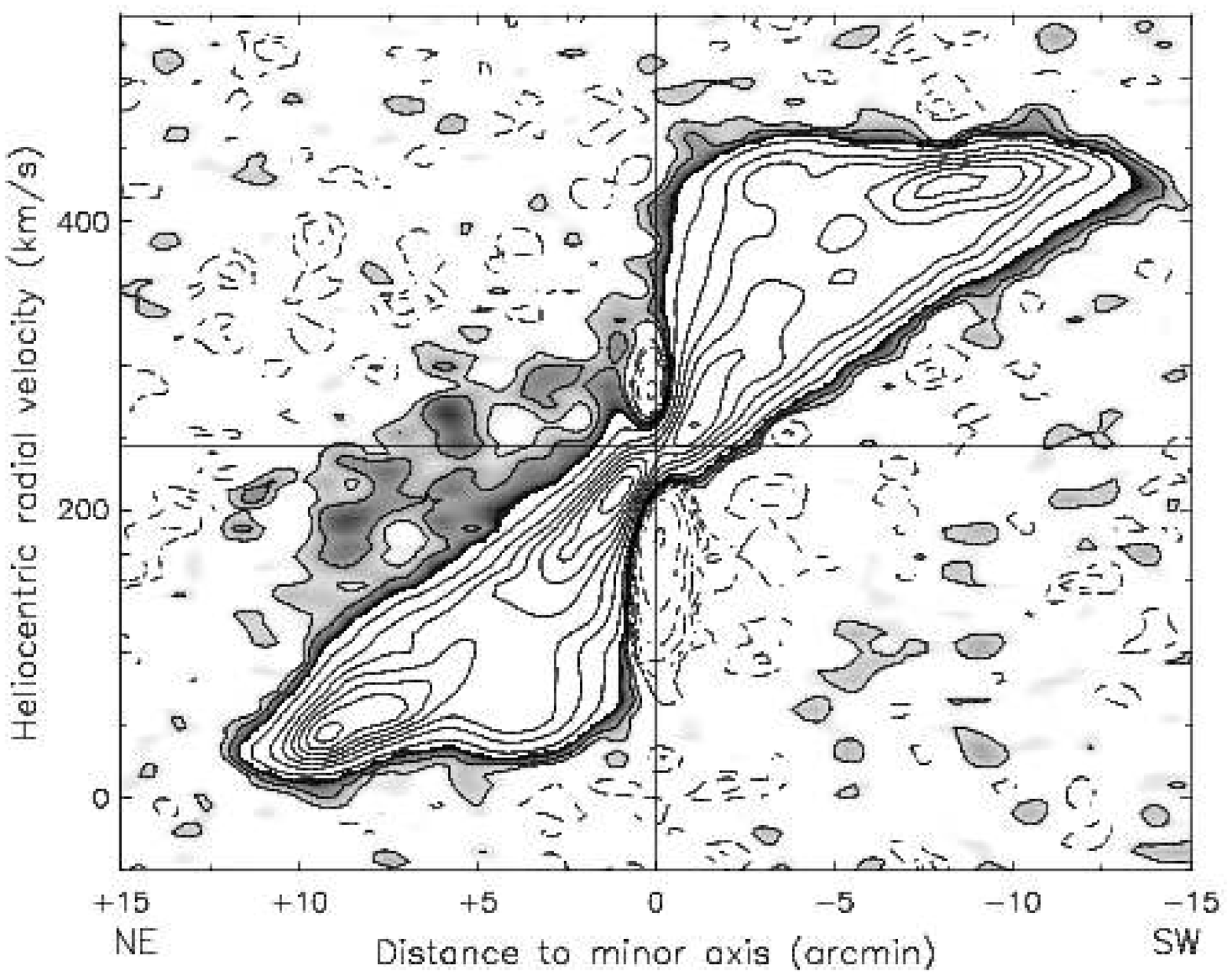}
\caption{Left panel: Total \HI\ map of NGC 253 on top of a DSS
image. Contours are at 0.18, 0.36, 0.72, 1.4, 2.9, 5.8, 12 and
23$\times 10^{20}$ cm$^{-2}$. Right panel: \HI\ position-velocity map
of \mbox{NGC 253} after strip-integrating in the direction parallel to
the minor axis. The halo emission is visible at the northeast side of
the galaxy, near the systemic velocity. Contours are at -4, -2, -1,
-.5 -.25 -.125, .125, .25, .5, 1, 2, 3, 4, 5, 6 and 7 Jy/beam.}
\end{figure}
Our new, deep \HI\ observations of \mbox{NGC 253} \citep{boo05} show
that there is indeed \HI\ above the disk of this starburst galaxy. The
total \HI\ distribution (Fig. 4, left panel) shows plumes of \HI\
extending 8$^{\prime}$ to 11$^{\prime}$ (9 to 12 kpc in projection)
from the midplane. The plumes straddle the extra-planar X-ray emission
\citep{pie00,str02}, and the diffuse H$\alpha$ emission
\citep{str02}.\\ The plumes are part of a larger complex of anomalous
\HI. This gas shows up clearly in the right panel of Fig. 4 as faint
emission around the systemic velocity. This position-velocity diagram
showing the overall kinematics of \HI\ in \mbox{NGC 253} has been
obtained by strip-integrating in the direction parallel to the minor
axis. The anomalous halo gas is seen only at the northeastern side of
the galaxy.\\ We have separated the anomalous \HI\ from the normal,
rotating \HI\ disk and found that it amounts to $8\times10^{7}$
M$_{\odot}$ of gas. This is about 3\% of the total \HI\ mass of
\mbox{NGC 253} ($2.5\times10^{9}$ M$_{\odot}$).\\ To investigate the
characteristics of the anomalous gas, we have constructed models of
the 3-D structure and kinematics of \mbox{NGC 253} consisting of a
thin \HI\ disk and a thick disk with different kinematics. From this
we conclude that \mbox{NGC 253} has a thick disk with a large, central
cavity. This disk is only partly filled with \HI, mostly in the NE,
and rotates more slowly than the thin disk. The \HI\ plumes seen above
and below the disk may be filaments or wall-like extensions of the
ring to higher distances from the plane, brought up by the central
superwind.\\ It is likely that, as already suggested for the X-ray and
H$\alpha$, the origin of the extra-planar \HI\ is also related to the
central starburst and to the active star formation in the central
disk. Alternatively, we may be seeing the effects of a minor merger
and gas accretion.\\

\section{Concluding remarks}

\mbox{NGC 6946} and \mbox{NGC 253} clearly are two more examples of
galaxies with extra-planar \HI. Yet no consistent picture emerges and
a number of processes may be at work. In both cases the data suggest a
connection with the star formation in the disk. In \mbox{NGC 6946}
this is more moderate but present over the entire disk. In \mbox{NGC
253} this is more violent, especially in the inner region. It is not
clear why the amount of gas in the halo are so different in the two
galaxies. In the outer parts of \mbox{NGC 6946} we also find evidence
for gas accretion.\\

\begin{acknowledgements}
The Westerbork Synthesis Radio Telescope is operated by the
Netherlands Foundation for Radio Astronomy with financial support from
the Netherlands Foundation for the Advancement of Pure Research
(N.W.O). The Australia Telescope is funded by the Commonwealth of
Australia for operation as a National Facility managed by CSIRO.
\end{acknowledgements}


\begin{thebibliography}{}
\bibitem[Bland-Hawthorn, Freeman \& Quinn(1997)]{bla97}
Bland-Hawthorn, J., Freeman, K. C., Quinn, P. J. 1997, ApJ, 490, 143
\bibitem[Boomsma et al.(2005)]{boo05} Boomsma, R., Oosterloo, T., Fraternali, F., van der Hulst, J. M., Sancisi 2004, A\&A, accepted for publication
\bibitem[Bregman(1980)]{bre80} Bregman, J. N. 1980, ApJ, 236, 577
\bibitem[Carilli et al.(1992)]{car92} Carilli, C.L., Holdaway, M.A.,
Ho, P.T.P., De Pree, C.G. 1992, ApJ, 399, L59
\bibitem[Ferguson et al.(1998)]{fer98} Ferguson, A. M. N., Wyse, R. F. G., Gallagher, J. S., Hunter, D. A. 1998, ApJ, 506, 19
\bibitem[Fraternali et al.(2001)]{fra01} Fraternali, F., Oosterloo,
T. A., Sancisi, R., van Moorsel, G. 2001, ApJ, 562, L47
\bibitem[Fraternali et al.(2003)]{fra03} Fraternali, F. 2004, IAUS,
217, 44
\bibitem[Heckman, Armus \& Miley(1990)]{hec90} Heckman, T. M., Armus,
L., \& Miley, G. K. 1990, ApJ Suppl., 74, 833
\bibitem[Hoopes et al.(1996)]{hoo96} Hoopes, C. G., Walterbos,
R. A. M., \& Greenawalt, B.E. 1996, AJ, 112, 1429
\bibitem[Van der Hulst \& Sancisi(1988)]{vdh88} van der Hulst, J. M., Sancisi, R. 1988, AJ, 95, 1354
\bibitem[Ibata (1999)]{iba99} Ibata, R. A. 1999, IAUS, 186, 39 
\bibitem[Jerjen et al.(1998)]{jer98} Jerjen, H., Freeman, K.C., Binggeli, B. 1998, AJ, 116, 2873
\bibitem[Kamphuis \& Sancisi(1993)]{kam93} Kamphuis, J., Sancisi, R. 1993, A\&A, 273, 31 
\bibitem[Karachentsev et al.(2003)]{kar03} Karachentsev, I.D., Grebel,
E.K., Sharina, M.E., Dolphin, A.E., Geisler, D., Guhathakurta, P.,
Hodge, P.W., Karachentseva, V.E., Sarajedini, A., \& Seitzer, P. 2003,
A\&A, 404, 93
\bibitem[Koribalski et al.(1995)]{kor95} Koribalski, B., Whiteoak,
J.B., \& Houghton, S. 1995, PASA, 12, 20
\bibitem[Matthews \& Wood(2003)]{mat03} Matthews, L. D. \& Wood, K.  2003, ApJ,  593, 721
\bibitem[Pietsch et al.(2000)]{pie00} Pietsch, W., Vogler, A., Klein,
U., \& Zinnecker, H. 2000, A\&A, 360, 24
\bibitem[Schaap et al.(2000)]{sch00} Schaap, W. E., Sancisi, R.,
Swaters, R. A. 2000, A\&A, 356, L49
\bibitem[Strickland et al.(2002)]{str02} Strickland, D. K., Heckman
T. M., Weaver, K. A., Hoopes, C. G., Dahlem, M. 2002, ApJ, 568, 689
\bibitem[Swaters et al.(1997)]{swa97} Swaters, R. A., Sancisi, R., van
der Hulst, J. M. 1997, ApJ, 491, 140
\end{thebibliography}
\end{document}